\documentclass[12pt]{article}
\usepackage[dvips]{graphicx}
\newcommand{\be}{\begin{equation}}
\newcommand{\ba}{\begin{array}}
\newcommand{\bd}{\begin{displaymath}}
\newcommand{\ee}{\end{equation}}
\newcommand{\ea}{\end{array}}
\newcommand{\ed}{\end{displaymath}}

\def\lsim{\mathrel{\rlap{\lower4pt\hbox{\hskip1pt$\sim$}}
    \raise1pt\hbox{$<$}}}      
\def\gsim{\mathrel{\rlap{\lower4pt\hbox{\hskip1pt$\sim$}}
    \raise1pt\hbox{$>$}}}      

\def\frac#1#2{{#1\over #2}}
\def\Imag#1{\Im{\rm m}#1}

\def\ds {\displaystyle}

\def\g2{ GeV$^2$}

\def\ie{\hbox{\it i.e. }}

\def\eg{\hbox{\it e.g. }}

\def\etal{\hbox{\it et al. }}

\newcommand{\EPJC}[1]{Eur.\ Phys.\ J. C #1}

\newcommand{\PRD}[1]{Phys.\ Rev.\ D #1}

\begin{document}

\bigskip 

\begin{center}
\bigskip

{\Large {Space structure of hadrons in elastic scattering at high energy}}
\bigskip

{\bf 
P. Desgrolard~\footnote{  E-mail: desgrolard@ipnl.in2p3.fr}
}

\end{center}

\bigskip

\noindent
Institut de Physique Nucl\'eaire de Lyon, IN2P3-CNRS et
Universit\'e Claude Bernard, 43 boulevard du 11 novembre 1918, F-69622
Villeurbanne Cedex, France

\bigskip

{\bf Abstracts} 
  
We have confirmed and complemented previous conclusions from studies of
the impact parameter profiles, relative to $pp$ and $\bar pp$ scattering,
using a generalized eikonalized amplitude. In particular,
the transition from the grey to the black disk is followed, when the energy
increases beyond $\sim 3$ TeV and for the inelastic process, by a black ring 
surrounding the antishadow scattering mode at the center.
The ranges of hadronic forces responsible for the 
elastic and inelastic processes are estimated by means of
root-mean-squares of the impact parameter.
Predictions at the LHC energy are explicitly given.
The effect of eikonalization is discussed.
\section{Introduction}
Since the pionnering work of Amaldi and Schubert, more than twenty years ago, 
the impact parameter (or $b$)-representation has a long history in 
the analysis of high-energy $pp$ and $\bar pp$ elastic scattering 
(\eg see~\cite{as} and references therein). 

Beyond the discussion of the elastic profile and overlap 
function (\eg~\cite{hv}), of the shadow and antishadow 
scattering mode~\cite{tt1}, extrapolations have also been
investigated~\cite{tt2} at the forthcoming facilities of the CERN 
Large Hadron Collider (LHC).

Aside from these explorations, the following mean-squares of impact parameter 
have not enjoyed much popularity since their introduction~\cite{ett}
\be
<b^2(s)>_{tot}\ =\ \ds {{\int_0^\infty b\ db\cdot b^2\ \Imag{H(s,b)}
\over\int_0^\infty b\ db\cdot\ \Imag{H(s,b)}}} \ ,
\ee
\be
<b^2(s)>_{el}\ =\ \ds {{\int_0^\infty b\ db\cdot b^2\ |H(s,b)|^2
\over\int_0^\infty b\ db\cdot\ |H(s,b)|^2}}\ ,
\ee
\be
<b^2(s)>_{in}\ =\ \ds {{\int_0^\infty b\ db\cdot b^2\ G_{in}(s,b)
\over\int_0^\infty b\ db\cdot\  G_{in}(s,b)}} \ ,
\ee
$H(s,b)$ is the elastic scattering amplitude in the
$b$-representation (or profile function) at a center of mass energy $\sqrt{s}$ 
and $G_{in}(s,b)$ the inelastic overlap function, sum of all 
inelastic channels contributions.
The reason for this lack of interest up to recently~\cite{kundrat0}
is probably due to the fact that it is
often convenient and sufficient to discuss the three quantities 
$\Imag{H(s,b)},\ |H(s,b)|^2,\ G_{in}(s,b)$ versus $s$ and $b$.

From a fundamental point of view, it is certainly 
worth to discuss also the corresponding "root-mean-squares" of the impact
parameters ($b$-rms)
\be\label{brms}
\sqrt{<b^2(s)>_{tot}}\ ,\ \sqrt{<b^2(s)>_{el}}\ ,\ \sqrt{<b^2(s)>_{in}}\ ,
\ee
in terms of characteristics of the averaged ranges of
hadronic forces responsible for elastic and inelastic collision processes.
The estimation of the $b$-rms could provide a new light 
on a simple physical picture at very high energy, in particular at the LHC, 
where the antishadow scattering mode is expected to be a potential 
discovery (see~\cite{tt2} and references therein).

It is the aim of this note to complement a traditional study of the
impact parameter elastic and inelastic profiles with a description of the  
space structure of hadrons in elastic scattering at high energy,
especially of protons at LHC.

\section { Choice of an amplitude}
We note the definitions (1-3) are written
in accordance with the notations used in the unitarity equation
\be 
\Imag{H(s,b)} = |H(s,b)|^2 + G_{in} (s,b)\ ,
\ee
which integrated over $b$, running from zero to the infinity, gives simply the
summation of the cross-sections~:  
$
\sigma_{tot}(s)\ =\ \sigma_{el}(s)\ +\ \sigma_{in}(s)\ .
$
 
In order to estimate $H(s,b)$, $G_{in} (s,b)$ and the $b$-rms, we have chosen 
the dipole version of the generalized eikonalized (GE) scattering amplitude 
found in~\cite{poes} for both $pp$ and $\bar pp$ elastic scattering processes.
It concerns a generalization of the ordinary eikonal method, constrained by
unitarity~\cite{eikun}, taking into account all intermediate multiparticle 
states involving the crossing-even and crossing-odd combinations of Reggeons. 
The formalism involves three parameters, $\lambda_0,\lambda_+,\lambda_-$,
corresponding to the three possible intermediate states.
At the Born level the amplitude involves
two secondary Reggeons, constructed from a simple Regge pole in the angular 
momentum $j$-plane, a Pomeron-"dipole", linear combination of a simple with a
double pole and finally an Oddderon-dipole including a factor for damping at 
$t=0$. 

The reasons for such a choice are:
\begin{itemize}
\item
it is a recently published amplitude, implying  the most general 
(to our knowledge) treatment of the eikonalization procedure;
\item
it leads to a very satisfying fit 
up to the largest explored 4-momentum transfer squared ($q^2=-t=14$ \g2 , 
neigbouring with the limit of validity of the Regge theory at $\sqrt{s}=27$ GeV)
and up to the Tevatron energy (1800 GeV)~\footnote{
The $pp$ cosmic rays data up to 40 TeV were of no help in the fit because of 
their too large errors, but they are also reproduced by the model. The 
prediction of the total cross-sectiom at the LHC 
$\sigma_{tot}=107 $ mb agrees with most common predictions in the literature
(\eg, see~\cite{bhs}). }.
\end {itemize} 
We add that the effect of eikonalizing the Born amplitude, on the profiles and 
on the $b$-rms and related quantities, may be easily 
evaluated. 
We notice that the present nuclear amplitude~\cite{poes}, that we use, involves
an eikonal unitarization while in~\cite{tt2,djs}, that we discuss, a $U$-matrix 
unitarization has been assumed.

In our opinion, the choice of an amplitude, respecting the Froissart-Martin 
bound and accounting data available in the widest range of high energies
and transfers is crucial to get confidence on
predictive power outside the fitted sets of data. The CERN 
LHC at $\sqrt{s}=14$ TeV
- and also the BNL Relativistic Heavy Ion Collider (RHIC) at a lower energy 
$\sim 500$ GeV - is probably the most interesting case to discuss 
presently (remember \eg the accepted project~\cite{totem} plans LHC 
measurements up to $|t|=10$ \g2). 

\section { Results and discussion}
Firstly, in order to test and generalize the model independence of the main 
conclusions 
already stated in~\cite{djs}, we performed a similar study, but with the Dipole 
GE amplitude of~\cite{poes}.
We reproduced similar results concerning
$\Imag{H(s,b)}$ and $G_{in} (s,b)$, versus $b$, up to the highest energies. 
\begin{enumerate}
\item 
An example is given Table 1 for the
central opacity of the nucleon $\Imag{H(s,0)}$. The agreement is good with 
existing experimental values. Prediction is also given for the LHC (where the
black disk limit is overtaken)
 
\begin{table}[ht]
\label{tab1}
\begin{center}
\begin{tabular}{|c||c|c|c|c|}
\hline
Energy (GeV)& 53    &546               &     1800         & 14000 \cr
            &(ISR)  &(S$\bar pp$S)     &(Tevatron)        &(LHC)  \cr 
\hline
Exp.        & 0.360 & 0.420$\pm$ 0.004 & 0.492$\pm$ 0.008 & ----- \cr  
Calc.       & 0.368 & 0.435            & 0.488            & 0.592 \cr
\hline
\end{tabular}
\end{center}
\caption{Central opacity of the nucleon,$\Imag{H(s,0)}$, calculated with the
Dipole GE amplitude  of~\cite{poes} and compared  to available experimental 
values~\cite{con} .
}
\end{table}
\item
In qualitative agreement with~\cite{tt2,djs}, we confirm that $\Imag{H(s,b)}$, 
respecting the unitarity limit 1, remains central all the way, exceeding the 
black disk limit 1/2 above the threshold calculated value 
$$
\sqrt{s}\sim 3\ {\rm TeV}
$$  
and then $G_{in} (s,b)$ undergoes a transition~: from central it becomes 
peripheral. In other terms we find that, beyond this threshold, its maximum 
(= 1/4) is shifted from $b=0$ towards increasing $b$.
The "interaction radius" ($R=R(s)$), where the maximal 
absorption takes place is \eg $R \sim .6$ fm at 14 TeV (see Fig.1) and would
smoothly becomes $R \sim 1.2$ fm at 1000 TeV.
We also find 
a hollow appearing at the center in $G_{in} (s,b)$ becoming deeper and deeper
when $s$ increases: \eg 0.23 at 14 TeV (Fig.1), 0.15 at 1000 TeV.
\item  It is a consequence of the chosen model~\cite{poes,eikun} that
$H(s\to\infty,b)\to{i\over 2\lambda_+}$, with $\lambda_+=0.5$, respecting
the unitarity inequality $|H(s,b)|\leq 1$, dictates 
at the center $\Imag{H}(s\to\infty,b=0)\to 1$, $G_{in} (s\to\infty,b=0)\to 0$. 
\item
Thus, from our calculations, the appearance of the antishadow mode reveals 
at small values of $b$, in a natural way at a high energy,
in conformity with its original introduction~\cite{tt1} and further
studies (\eg~\cite{tt2,djs}). It is worth emphazing
that the antishadow scattering mode is predicted here using a hadron-hadron 
amplitude quite different from those used in other calculations. 
This enforces the model independence of such a prediction but of course only
experiments at very high energy could decide whether it exists or no. 
We found a calculated energy threshold for its apparition 
above the Tevatron, but below the LHC~\footnote{
It has been estimated 2 TeV in~\cite{tt2} and 6 TeV in ~\cite{djs}, to be 
compared to 3 TeV here.
}. 
\item
It is interesting to mention the effect of eikonalization 
by changing $H,G_{in}(s,b)$, once eikonalized and fitted, into
$h,g_{in}(s,b)$, at the Born level and non-refitted. Returning to the Born level 
brings forwards the antishadow mode characterics \ie minoration of the energy
threshold and majoration of the interaction radius. The drastic effect 
of eikonalization, well known on the currently fitted observables,
is clearly visible, at the LHC energy for example, on $\Imag{H}$ and $G_{in}$
(see Fig.1): the antishadow scattering  holds at $b\lsim 0.6$ fm  
after eikonalization and at $b\lsim 0.9$ fm
at the Born level (these values correspond to the maximum 1/4 taken by $G_{in}$
and to the crossing of the black disk limit 1/2 by $\Imag{H}$ at the LHC).
\end {enumerate} 

\bigskip

Secondly, in order to complete the preceding study, we have also calculated 
the three (total, elastic and
inelastic) root-mean-squares of the impact parameter, defined in (\ref{brms})
and interpreted as characteristics of the ranges of hadron forces
responsible for corresponding collisions.
There are plotted in Fig.2, versus $\sqrt{s}$, at 
the Born level and after the generalized eikonalization.

The main remarks suggested by these curves are in order
\begin{enumerate}
\item
A common gross characteristic of the $b$-rms is to have a mild dependence on 
the
process and on the energy beyond the ISR up to the LHC. Roughly speaking, they 
lie between 0.7 fermi and 1.2 fermi. 
\item
They show almost parallel and linear increases with the energy.
\item 
To be more precise
$\sqrt{<b^2(s)>_{tot}}$ goes from 0.9 to 1.2 fm; 
$\sqrt{<b^2(s)>_{el}} $ goes from 0.6 to 0.9 fm;
$\sqrt{<b^2(s)>_{in}} $ goes from 1.0 to 1.3 fm,
when the energy increases through four decades from $10$ GeV to $100$ TeV.
\item
Strictly speaking, the effect of eikonalization is poor on these quantities,
in contrast with such effect on current observables (\eg cross-sections) or on 
the elastic profile or on the inelastic overlap function, but in
agreement with their definitions as averages in
the $b$ representation which smoothen the differences.
The two first pairs of curves (total and elastic) yield almost parallel 
increases with the energy, the GE case being above the Born case, while the 
third one (inelastic) exhibits a Born $b$, close to GE $b$ and becoming greater
at high energy. In other terms the inelastic Born line crosses the GE line
in the TeV region,
 this crossing corresponds to the threshold value of the 
antishadow mode and recalls the importance of neglecting
the eikonalization on the inelastic function as discussed above.

\end {enumerate}

\noindent
The results shown in Fig.2 are in
agreement with the results from an analysis~\cite{kundrat} at 53 GeV and 
546 GeV, calculated in a central hypothesis, \ie with a non-existent degree 
of peripherality (see Table 2), 
in spite of strong differences in constructing the nuclear amplitude. 
Furthermore, we note, that the agreement still holds for $\sqrt{<b^2(s)>_{tot}}$
when a degree of 
peripherality is entered in the amplitude. This conforts us in the model
independence of the present results.

\begin{table}[ht]
\label{tab2}
\begin{center}
\begin{tabular}{|c||c|c|c|c||c|c|}
\hline
Energy  (GeV)   &\multicolumn{2}{|c}{53}      & \multicolumn{2}{|c||}{541}  
                & 1800    & 14000  \cr
\hline 
                &    P.W.   &  \cite{kundrat}  &   P.W    &  \cite{kundrat} 
		&    P.W  &  P.W   \cr         
\hline
$\sqrt{<b^2   >_{tot}}$ (fm)  & 1.00 & 1.03 & 1.04& 1.13 & 1.07 & 1.14  \cr  
$\sqrt{<b^2   >_{el}} $ (fm)  & 0.67 & 0.68 & 0.72& 0.76 & 0.76 & 0.83  \cr             
$\sqrt{<b^2   >_{in}} $ (fm)  & 1.06 & 1.09 & 1.11& 1.21 & 1.16 & 1.26  \cr
\hline
\end{tabular}
\end{center}
\caption{Total, elastic, inelastic root-mean-squares calculated with the 
GE amplitude  of~\cite{poes}, at four representative energies, in this
present work (P.W). The values of~\cite{kundrat} for the two lowest
energies, evaluated in a central hypothesis,
are also quoted.
}
\end{table}
\section {Conclusion}
In this note, we have used an amplitude~\cite{poes} fitted, after a 
generalized eikonalization process, reproducing all $pp$ and $\bar pp$ 
elastic scattering data. 
We have confirmed and completed conclusions of previous papers~\cite{tt2,djs} 
from a study of the impact parameter profiles for elastic and inelastic 
processes. They concern in particular the transition from the grey to the black
disk, which is expected to be followed by the inelastic overlap function and
when the energy increases beyond a threshold at $\sim 3$ TeV, by a black ring
surrounding the antishadow scattering mode at the center ($b=0$).

According to this scenario,
the old picture of a proton getting "BEL" (blacker, edgier, larger) in $pp$ and
$\bar pp$ collisions, when the energy increases, still holds for both elastic
and inelastic processes below the
threshold. Above (let us say in the region of LHC energies), 
the antishadow mode appears in the inelastic collisions for which the
proton behaves like a half-transparent core (a grey disk), with an outer shell 
(a black ring), mainly responsible for inelastic process. No echo is found on
the traditional behaviour of the proton in the elastic and total processes.
 
The ranges of hadronic forces  responsible for elastic and inelastic processes,
estimated by means of
root-mean-squares of the impact parameter are found respectively
$\sim 0.8$ fm and $\sim 1.3$ fm, at LHC energy. They are only weakly dependent
of the eikonalization, enforcing the interest of the Born approximation when
estimating such characteristics.

\bigskip

{\bf Acknowledgements.} I would like to thank V. Kundrat, S. Troshin,  
N. Tyurin for useful comments and E. Martynov for a critical reading of the
manuscript.

\vfill\eject
\vglue -3cm

\begin{figure}[t]
\label{unit}
\begin{center}
\includegraphics[scale=0.90]{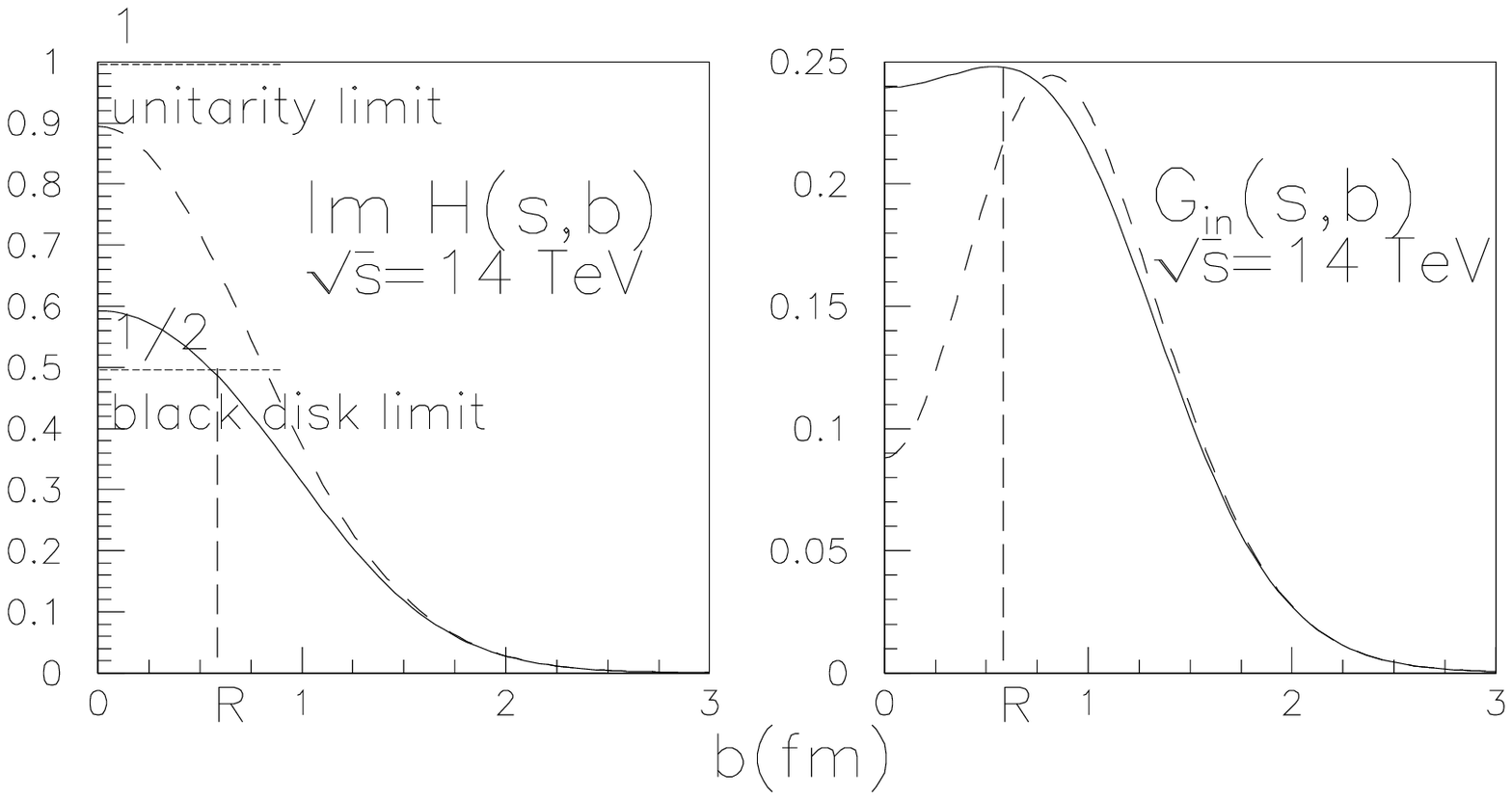}
\end{center}
\end{figure}

\vskip 2.0cm
\noindent
Figure 1. $\Imag{H(s,b)}$ and inelastic overlap function $G_{in} (s,b)$
versus the impact parameter $b$ calculated at the LHC energy
in the GE formalism, with the fitted
amplitude of~\cite{poes} for the $\bar pp$ elastic scattering (solid line). 
$R$ is the "interaction radius" (see the text).
A comparaison is shown with the -non-fitted- Born 
predictions (dashed line), proving the effect of eikonalization  which soffens
the antishadow characteristics.

\begin{figure}[h]
\label{bfig}
\begin{center}
\includegraphics[scale=0.87]{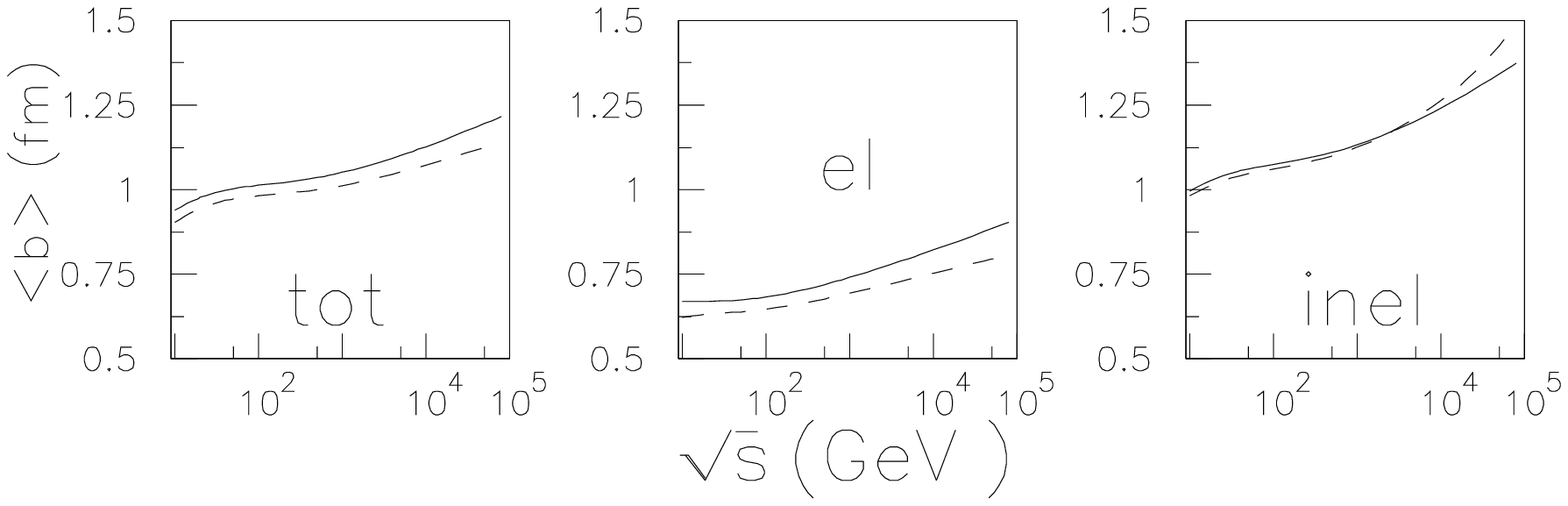}
\end{center}
\end{figure}

\vskip -0.8cm
\noindent
Figure 2. Total, elastic, inelastic root-mean-squares of impact parameter (5)
calculated in the GE formalism, with the fitted
amplitude of~\cite{poes} for the $\bar pp$ elastic scattering (solid line, the
$pp$ plots would be indiscernable). A comparaison is shown with the Born
predictions (dashed line), proving the small effect of eikonalization on these
quantities.

\end{document}